\documentclass[showpacs,twocolumn]{revtex4}
\usepackage{txfonts}
\usepackage{amssymb}

\usepackage{graphicx}
\usepackage{dcolumn}
\usepackage{bm}

\begin{document}

\title{\bigskip  Phase Compensation Enhancement of Photon Pair Entanglement Generated from Biexciton Decays in Quantum Dots }
\author{Zong-Quan Zhou $^{1,2}$ , Chuan-Feng Li$^{1,}\footnote{
email: cfli@ustc.edu.cn}$, Geng Chen$^1$, Jian-Shun\nolinebreak[4]
Tang$^1$, Yang Zou$^1$, Ming Gong$^1$ and Guang-Can Guo$^1$}
\affiliation{ ${}^1$Key Laboratory of Quantum Information,
University of Science and Technology of China, CAS, Hefei, 230026,
People's Republic of China
 \\ ${}^2$ School of Instrumentation and Opto-electronics Engineering, Hefei University of Technology, Hefei, 230009, People's Republic of China }
\date{\today }

\begin{abstract}
Exciton fine-structure splittings within quantum dots introduce
phase differences between the two biexciton decay paths that greatly
reduce the entanglement of photon pairs generated via biexciton
recombination. We analyze this problem in the frequency domain and
propose a practicable method to compensate the phase difference by
inserting a spatial light modulator, which substantially improves
the entanglement of the photon pairs without any loss.

\end{abstract}

\pacs{03.67.Bg 78.55.Cr 42.50.-p}
\maketitle

Entangled photon pairs play a crucial role in much of quantum
information processing \cite{Spe1998,A1991,C1992,C1993}. The most
widely used methods for generating entangled photon pairs involve
nonlinear optical processes, such as spontaneous parametric down
conversion (SPDC) \cite{White1998,cfli2006}. However, high
multi-photon probabilities and low quantum efficiencies associated
with SPDC pose serious limitations on their applications in quantum
information processing.

As an alternative, biexciton decays in single quantum dots (QDs)
have been proposed as good sources of ``on-demand" entangled photon
pairs \cite{Benson2000}. QDs also have the advantages of a mature
fabrication technology and ease of integration into larger
structures to make monolithic devices. However, ``which-path"
information provided by the fine-structure splitting (FSS) of the
intermediate exciton state destroys the entanglement of photon pairs
\cite{Santori2002}. To overcome this problem, the energy splitting
is tuned to near zero either by rapid thermal annealing
\cite{Young2005}, or optionally applying in-plane electric fields
\cite{Electric2007}, magnetic fields
\cite{Magnetic2006,Young2006Na}, uniaxial stresses \cite{Stress2006}
or light fields \cite{Light2009}. Such ``triggered" entangled photon
pair sources can also be engineered by simply selecting appropriate
QDs with small FSSs \cite{Hafenbrak2007}, by energy-resolved
post-selection \cite{Petroff2006}, and by using highly-symmetric,
site-controlled quantum dots grown in inverted pyramids
\cite{Mohan2010}.

In considering the photon emission distribution in the time domain,
the two-photon state created in a QD is \cite{Young2007}
\begin{eqnarray}
 \Psi(t)=(\sqrt{\frac{1}{\tau}e^{\frac{-t}{\tau}}}
H_{XX}H_{X}+\sqrt{\frac{1}{\tau}e^{\frac{-t}{\tau}}}e^{iSt/\hbar}
V_{XX}V_{X})/\sqrt{2},
\end{eqnarray}
where $S$ denotes the FSS energy, $t$ is the time delay between the
first (biexciton) and the second (exciton) photon emission events,
$\frac{1}{\tau}e^{\frac{-t}{\tau}}$ is the exciton photon emission
probability distribution, and $\tau$  is the exciton lifetime. Thus
time integration will reduce the overall degree of entanglement, and
even lead to classically correlated states \cite{Young2008}.

In this Report, we analyze this problem in the frequency domain and
propose an optical arrangement to compensate the phase difference.
Given the maximally entangled state $
(H_{XX}H_{X}+V_{XX}V_{X})/\sqrt{2} $, the fidelity is greatly
improved and in our method is accompanied without photon losses,
thus surpassing previous schemes that apply timing gates
\cite{Young2008} and employ energy-resolved post selection
\cite{Petroff2006}.

By Fourier Transformation, we can re-express the two-photon state in
the frequency domain
\begin{eqnarray}\Psi(\omega)=(f_{_H}(\omega)e^{i\varphi_{_H}}H_{XX}H_{X}+
f_{_V}(\omega)e^{i\varphi_{_V}}V_{XX}V_{X})/\sqrt{2},
\end{eqnarray}
 with
 $f_{_H}(\omega)=\{{2\pi\tau[1/(2\tau)^{2}+1/\omega^{2}]}\}^{-1/2},\ \varphi_{_H}=\tan ^{-1}(-2\omega\tau),
\
f_{_V}(\omega)=\{{2\pi\tau[1/(2\tau)^{2}+1/(S/\hbar-\omega)^{2}]}\}^{-1/2},\
 \varphi_{_V}=\tan ^{-1}[2\tau(S/\hbar-\omega)]. $

The polarization density matrix is given by
 \begin{eqnarray}
 \rho=\frac{1}{2}\left(
              \begin{array}{cccc}
                1 & 0 & 0 & \alpha \\
                0 & 0 & 0 & 0 \\
                0 & 0 & 0 & 0 \\
                \alpha^\ast & 0 & 0 & 1 \\
              \end{array}
            \right)
           \end{eqnarray}
with $\alpha=\int f_{H}(\omega)f_{V}(\omega) e^{i\varphi} d\omega$
and $\varphi=\varphi_{_V}-\varphi_{_H}$ is the phase difference. The
fidelity with Bell state is
\begin{eqnarray}
f=\frac{1}{2}[1+\int f_{_H}(\omega)f_{_V}(\omega)\cos\varphi
d\omega].
 \end{eqnarray}

\begin{figure}[tbph]
\begin{center}

\includegraphics[width= 3.3in]{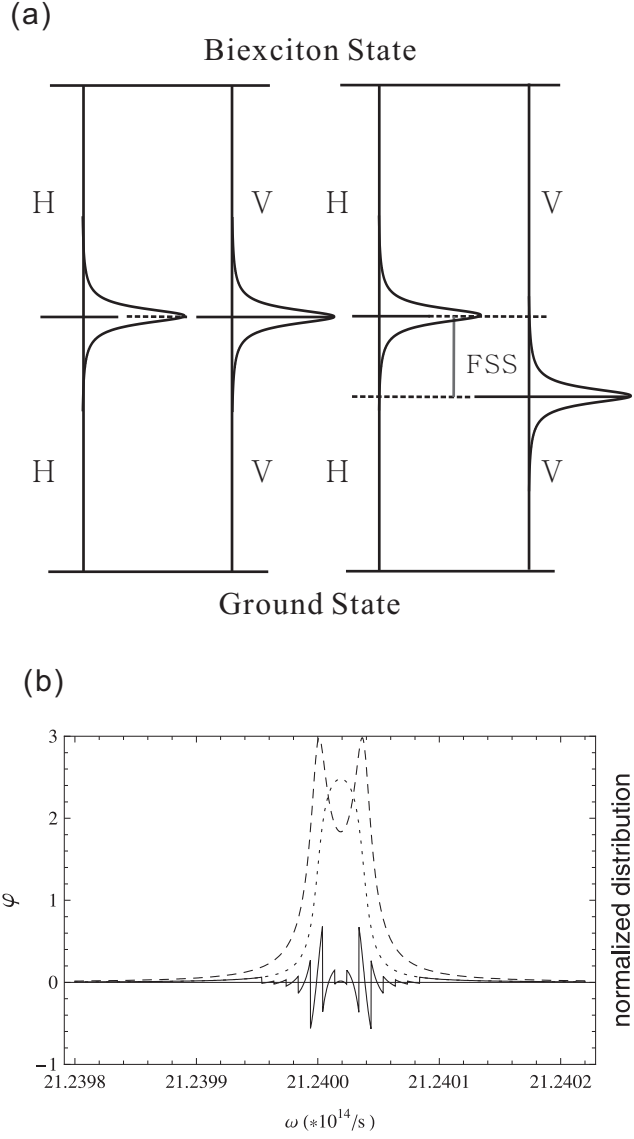}
\end{center}
\caption{ (a) The level diagram of the radiative decay of the
biexciton state. In the left panel, the FSS energy $S$ of the QD is
zero, while in the right $S$ is nonzero. (b) A composite graph of
the normalized $f_{_H}(\omega)f_{_V}(\omega)$ distribution (dashed
line) and the phase difference between H and V polarization (dotted
line) as a function of angular frequency $\omega$ with $S$ of $2.5\
$meV. The solid line shows the phase after compensation as discussed
in the text.}
\end{figure}

FSS limits the degree of entanglement in two ways, as illustrated in
Figure 1. First, the phase difference between H- and V-polarized
photons reduces the fidelity after time integration. Second, the
overlap between their photon frequency distributions decreases as
the FSS energy increases. We concentrate first on the phase
difference.

Phase compensation is generally difficult to realize in the time
domain, because it requires an accurate phase delay
($\sim$$St/\hbar$) rapidly varying with time. As shown in Fig. 1(b),
the phase distribution is clearly a non-monotonic function of
frequency, so it is also impossible to realize this compensation by
simply using a dispersive element.  The proposed experimental setup
is shown in Fig. 2. The light from the QDs should initially be
collimated and focused. For widely-used self-assembled QDs, the
separation between biexciton (XX) and exciton (X) emission lines is
generally several \emph{m}eV because of the biexciton binding energy
\cite{Michio2006}, that then enables the use of a dichroic mirror
(DM) to separate them. The emitted XX photon goes directly to a
single photon detector (SPD). The X photon enters a polarization
beam splitter (PBS), which reflects the V polarization and transmits
the H polarization. The two parallel gratings distribute the photons
in a spatial mode depending on their wavelengths. The diffraction
angle $\theta$ is determined by $d\sin\theta-d\sin i=\lambda$, where
$i$ is the incident angle and $\lambda$ is the photon's wavelength.
The last mirror reflects the photon back. The quarter-wave plate
(QWP) is $22.5^{\circ}$ placed. Passing through the QWP twice
changes H (V) polarization to V (H) polarization. The half-wave
plate (HWP) and the polarizer placed before the detectors are used
to choose the polarization state for coincidence detection. All
gratings, incident angles, and optical paths are identical in both
arms of H and V polarization. As shown in Fig. 1(b), the phase
difference is driven close to zero after which is divided in many
small steps. Here each step corresponds to an angular frequency
bandwidth of $\Delta\omega\doteq 1\times10^{10} /\emph{s} $. Some
parameters have been chosen, the vertical distance between the two
parallel gratings is 0.29 meters, the gratings' constant $d=1.1\
\mu$m, and the incident angles satisfy $\sin i$ = 0.18. The angular
frequency $\omega=2.124\times10^{15} /{s} $, corresponds to a
wavelength of 0.887 $\mu$m. We choose the length of each step in
front of the mirror as $\Delta= 20\ \mu$m, which is much larger than
the photon wavelengths.

\begin{figure}[tbph]
\begin{center}
\includegraphics[width= 3.3in]{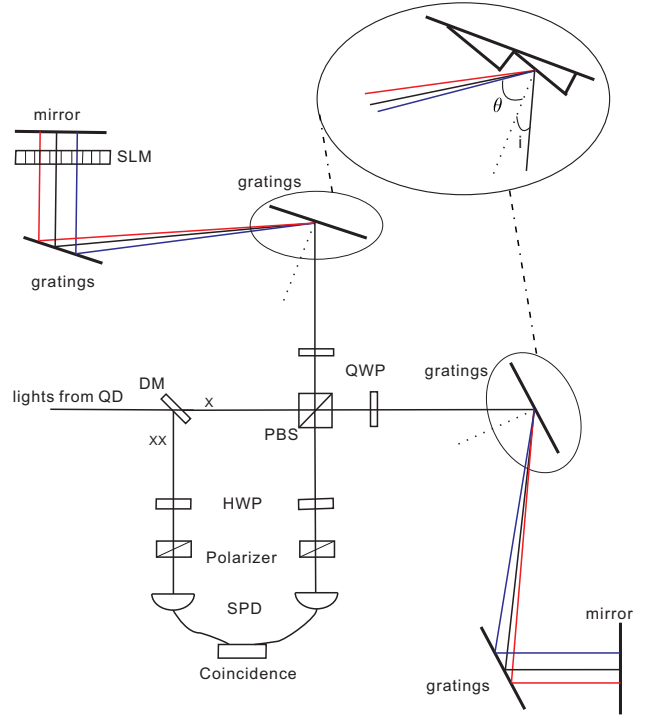}
\end{center}
\caption{ Experimental setup for entanglement phase compensation. X
and XX photons are separated by the DM. The XX photons go directly
to SPD, while the X photons are separated into two arms by PBS. To
realize the phase compensation, a SLM is inserted in one of the
paths.}
\end{figure}

Based on equation (4), the variation fidelity with FSS energy is
shown in Fig. 3, where the dotted line corresponds to the phase
difference $\varphi$ as mentioned above, and produces exactly the
same results as obtained in the time domain \cite{Young2007}. The
solid line corresponds to the ideal case without phase difference
(i.e, $\varphi=0$). Here the QD exciton lifetime $\tau$ is set to
0.77 ns which is consistent with experimental observations
\cite{Young2007}.

Several methods can be used to realize this sectionalized
compensation spatially, such as an optical coating with varying
thickness, a medium of varying refractive index, or Fiber Bragg
Grating (FBG) which is the standard dispersion compensation
techniques used in optical fiber communications \cite{K1994}.
However the more appropriate method would be to use a phase-only
spatial light modulator (SLM), which can change the phase delay
distribution spatially pixel by pixel with an electric signal. This
is important in this scheme as different QDs have different FSS
energies, and hence their phase distributions differ. The advantages
is that while other methods may require completely new fabrication
to adapt to certain QDs, here with a SLM, different QDs just require
changes in the electric signal for each pixel of the SLM. The phase
range shown in Fig. 1(b) can never be larger than $\pi$, therefore
the phase compensation can be easily realized with a SLM. Moreover,
SLMs with pixel resolution of 20 $\mu$m are commercially available,
and even 5 $\mu$m resolution can be achieved. A higher resolution
will give finer compensations, and thus obtain the fidelity closer
to the ideal altough at the expense of photon loss due to
diffraction. With 20 $\mu$m length steps, corresponding to the
bandwidth $\Delta\omega\doteq 1\times10^{10} / s $, the result is
already very close to the ideal case as shown in Fig. 3. The dashed
line gives the result after phase compensation and the solid line
shows the ideal case when $\varphi=0$. Furthermore, for $\Delta=20\
\mu m$, diffraction effects are negligible. The photon loss caused
by diffraction can be estimated as $\lambda/\Delta$, which is
vanishingly small in this case.

\begin{figure}[tbph]
\begin{center}
\includegraphics[width= 3.3in]{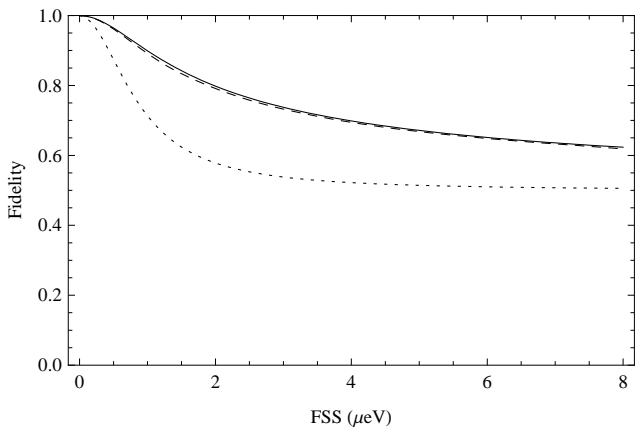}
\end{center}
\caption{ The variation of fidelity with FSS energy. The dotted line
is without phase compensation. The solid line shows the ideal case
without any phase difference while the dashed line gives the results
with phase compensation as mentioned in the text. }
\end{figure}

As shown in Fig. 3, with a FSS energy $S$ of 2.5 $\mu$eV, the
fidelity increases from 0.553 to 0.764 after phase compensation.
Even with $S$ of 3.8 $\mu$ev, the fidelity is still over 0.7 after
phase compensation. We have noted that in a published report that
with $S$ of 2.5 $\mu$eV obtained by applying a timing gate, the
fidelity increases from 0.46 with a gate width of 2 ns to 0.73 with
a gate width of 49 ps \cite{Young2008}. A simple calculation gives a
theoretical collective efficiency of 0.925 with the 2 ns gate, but
rapidly declines down to 0.061 with the 49 ps gate. It is obvious
that to get higher fidelity more photons have to be rejected by the
timing gate. In contrast, we improve the fidelity by a factor of
0.21 without any photon loss theoretically. Considering the
practical performance of the gratings (efficiency $\sim$ 90\%) and
the SLM (efficiency $\sim$ 95\% ), we estimate the experimental
efficiency of 62\% in performing phase compensation. The bare
postselection in energy \cite{Petroff2006} is even more wasteful
than applying a timing gate \cite{Young2008}, since one has there to
select a small fraction of photons with overlapping frequencies, and
further ensure that the phase difference does not change much in the
selected frequency band.

Even after phase compensation, the fidelity unfortunately cannot
attain unity, because it is limited by the photons outside the
overlapping part of the frequency. Further improvements in fidelity
can be achieved by rejecting these photons. For example, an even
better performance than that of the ideal case shown with solid line
in Fig. 3, can be achieved in cooperation energy-resolved
postselection as reported in Ref. \cite{Petroff2006}. With phase
compensation, this postselection can be more efficient and result in
a much enhanced performance in fidelity. To illustrate the point,
let S=2 $\mu$eV, if the angular frequency bandwidth is set at
$\{2.1240006\times10^{15} /\emph{s},\ 2.1240024\times10^{15}
/\emph{s}\}$, then after phase compensation, the fidelity increases
from 0.578 to 0.9 with a theoretical efficiency of 0.2 (still much
higher than that obtained by applying a timing gate). Our result
means that if a relative low efficiency can be tolerated, even a
non-zero FSS is acceptable and required no magnetic or other fields
to tune the FSS to zero, thus greatly simplifying the experimental
setup. As reported previously \cite{CFLi2008}, calculations reveal
that InAs/InP QDs offer smaller FSSs with only a little flux around
zero for individual QDs. Very recently, Mohan et al.
\cite{Mohan2010} reported that highly symmetric, site-controlled
quantum dots show FSS energies of several $\mu$eV. Utilizing these
types of QDs, our scheme may lead to a practical entangled photon
pairs source that is efficient and easy to control.

Another advantage of this setup is that it is easy to control the
output phase, an aspect that is highly desirable in a multitude of
contexts in quantum information processings. This is achieved simply
by either allowing the SLM to introduce a constant delay, or
changing the optical path in one of the arms.

It should be noticed that we have not considered the effect of spin
flipping and background light here, which in practice may slightly
degrade the experimental results. Further studies would need to
include the evolution of these effects and others, and try to solve
them in the frequency domain.

To summarize, we have analyzed the degradation in entanglement of
photon pairs emitted from QDs with non-zero FSS, and proposed a
phase compensation scheme with the insertion of a SLM to greatly
enhance entanglement gaining theoretical efficiencies approaching
100\%. An even better performance in fidelity can be achieved in
cooperation with frequency post-selection.

This work was supported by National Fundamental Research Program,
National Natural Science Foundation of China (Grant No. 60621064,
10874162, and 10734060).

\end{document}